# Optical control of spatially localized red blood cell activity by holographic tweezing

Niccolò Caselli[1,2,*], Mario García-Verdugo[1], Macarena Calero[1,2], Natalia Hernando-Ospina[1,2], José A. Santiago[3], Diego Herráez-Aguilar[4] and Francisco Monroy[1,2,#]

[1] Departamento de Química Física, Universidad Complutense de Madrid, Ciudad Universitaria s/n, 28040 Madrid, Spain
[2] Translational Biophysics, Instituto de Investigación Sanitaria Hospital Doce de Octubre, 28041 Madrid, Spain
[3] Departamento de Matemáticas Aplicadas y Sistemas, Universidad Autónoma Metropolitana Cuajimalpa, Vasco de Quiroga 4871, 05348 Ciudad de México, México.
[4] Instituto de Investigaciones Biosanitarias, Universidad Francisco de Vitoria, Ctra. Pozuelo-Majadahonda, Pozuelo de Alarcón, Madrid, Spain

**Email:** *ncaselli@ucm.es; #monroy@ucm.es

**Abstract**

Red blood cells possess unique biomechanical ability to squeeze through capillaries smaller than their size to enable gas and ion exchange. A key signature of their active biomechanics is the out-of-equilibrium fluctuation of the plasma membrane, also known as flickering motion. This active flickering is driven by motor proteins that connect the forces between the spectrin skeleton and the lipid bilayer. However, studying flickering motions in living red blood cells is challenging without altering their physical properties. Here, we implemented a holographic optical tweezer that sculpted a laser beam to create a force field distributed directly along the membrane equatorial contour. We show heterogeneous membrane flickering activity driven by membrane kickers in free-standing cells. Then we inhibited the active kickers by optical forces under minimal invasion, thus benchmarking the active motion against thermal fluctuations. Our work paves the way for optical control of biophysical forces, providing touchless strategies for mechanotransduction in living cells.

**Introduction**

Exploring cellular mechanical properties and the intricate mechanotransduction-metabolism interplay provides a deeper understanding of physiological functions and potential pathologies [1,2]. Mapping and inducing mechanical forces in living cells is of the utmost relevance not only for basic studies in experimental cell physics, but also in developing minimally invasive therapy e.g., for cell-machine interfacing [3], tissue manipulation and regeneration [4,5], osteogenic differentiation [6], and microsurgery [7]. Forces at the subcellular scale can be induced by atomic force microscopy [8], traction force microscopy [9], magneto-optical traps [10], and optical tweezers [11–14]. As living cells show minimal light absorption or scattering in the near-infrared, optical tweezing has been used to trap intracellular elements without phototoxicity [15]. This approach enabled *in vivo* micromanipulation using force-mediator beads [16,17] or direct bead-free trapping [18–20], establishing a suitable platform for cell microrheology [21]. The introduction of a hologram structuring coherent beam, which controls both the amplitude and phase of the optical field, gave rise to optical tweezers with multiple spatially-extended trapping [22], and even transfer of orbital angular momentum of light into trapped objects [23,24]. The notable advantage of holographic tweezers lies in their ease of integration into microscopy platforms by means of diffractive elements such as spatial light modulators (SLMs) [25–27]. In this work, we present a holographic tweezing device able to directly manipulate living cells inducing a trapping potential with a spatial distribution dictated by the cellular shape. This approach allowed us to regulate the mechanical behavior of red blood cells (RBCs), which are biological archetypes for testing single-cell biomechanics due to their remarkable activity fueled under glycolytic production of ATP[28,29]. RBCs, also known as erythrocytes, are autonomous anucleate cells with a specialized plasma membrane associated with a highly dynamic and actively flexible cytoskeleton that enables large deformations and adaptation to the physiological discocyte shape [30]. Wheater observed at the equatorial rim, living RBCs exhibit highly dynamic membrane undulations even when outside the vasculature [31], an active phenomenon known as flickering [32]. This active flickering serves is a marker of ATP-dependent metabolic activity in the plasma membrane [33,34]. In ATP-deficient erythrocytes the membrane becomes stiffer under inhibition of its glycolytic energy capacity and they undergo abnormal cell shapes (stomatocyte, echinocytes, etc.) as supported by a rigidified skeleton [30]. At the molecular level, the mechanobiological active flickering of RBCs is thought to emerge from underlying biochemical processes settled in the plasma membrane and in its connection to the cytoskeleton formed by a quasi-hexagonal network composed of spectrin tetramers and short actine filaments[35]. Arguably, active flickering could also involve membrane ion transporters, phosphoinositide lipids, ankyrin tethers, and phosphorylation of protein such as band 4.1 R-based macromolecular complexes that are the junctions between flexible membrane and active cytoskeleton[36,37]. The long filaments of non-muscular myosin anchored to the cytoskeleton via actin, although present in relatively small quantities within RBCs, have been also suggested as force effectors inducing energy-dependent flickering due to their extension and contraction activities[38].

Here, we prove the dynamically active nature of the flickering motions in RBCs by sculpturing trapping force landscapes directly on the equatorial membrane rim of single RBCs by employing holographic optical tweezers. Spatial maps of flickering activity were evaluated both for free-standing and optically trapped RBCs. In the freely flickering cells, we found heterogeneous active hot spots, i.e., performing large membrane deformations, consequently identified as membrane kickers. We hampered these membrane kickers by inducing an external force landscape by means of holographic tweezers under minimal invasiveness and negligible phototoxicity. Our proof-of-concept approach evidences a direct physical intervention on living cell membranes, thus envisaging disruptive applications in mechanobiology and micro-medicine.

**Results**

**Holographic optical tweezers**

To simultaneously achieve optical trapping and fast imaging of living cells, we implemented a home-made inverted microscope in which a laser coupled to a spatial light modulator (SLM) generates an extended optical tweezers [39]. Figure 1a depicts the optical platform (see Methods for details). Briefly, a near-infrared laser beam illuminates a reflective SLM, which is a digitally controlled liquid-crystal screen capable of changing the phase of reflected light. This technique can generate holograms with requested intensity distribution $I(x,y)$ in the reciprocal Fourier plane of the SLM [40–42]. Therefore, by focusing the sculptured laser beam on the sample plane ($z=0$), we were able to induce and sculpture locally optical forces due to the intensity gradient $\vec{F} \propto \vec{\nabla}I$. We chose stationary holograms $I(x,y)$ that reproduce the RBC membrane contour $R(x,y,z=0)$ observed in the equatorial plane. RBCs can be directly trapped by optical tweezing forces because their refractive index is larger in the near-infrared ($n_{RBC} \approx 1.4$), than the surrounding cell culture medium used for *in vitro* studies ($n_0 \approx 1.3$), giving a refractive index difference $\Delta n = n_{RBC} - n_0 \approx 0.1$ [33,43]. Therefore, optical trapping potentials $\mathcal{E}(x,y,z=0) \propto \Delta n\, I(x,y)$ can be exerted on the whole RBC membrane, which possesses larger permittivity than the surrounding medium (see Fig. 1b-c). To control the optical trapping landscapes, the equatorial RBC membrane shape was feedbacked to the SLM to achieve the hologram with intensity distribution $I(x,y) = R(x,y)$ (see Supporting Information Fig. S1). In polar coordinates, the trapping forces are locally deployed by the intensity distribution $I(r,\theta) \equiv I_0 \delta[r(\theta) - R(\theta)]$, where $I_0$ is the laser intensity and $\delta$ the Dirac's function that fixes the hologram at the equatorial rim $R(\theta)$. This feedback regulation allows us to control the RBC flickering activity by holographic optical tweezers.

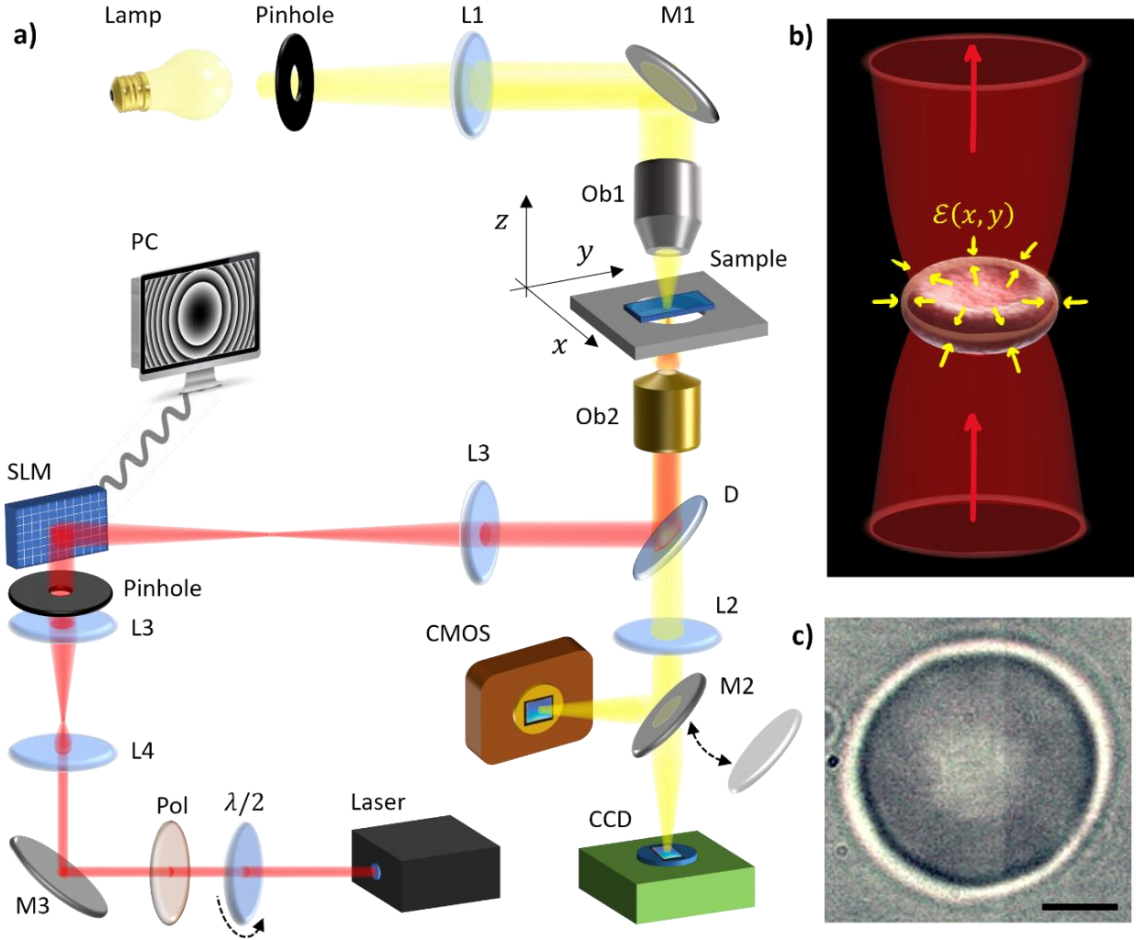

**Figure 1. Holographic optical tweezers setup. a)** Schematics of the holographic optical tweezers coupled to an inverted microscope. SLM is the spatial light modulator, $M_i$ mirrors, $L_i$ lenses, $\lambda/2$ half-wave plate, Pol linear polarizer, D dichroic mirror, Ob1 condenser objective, Ob2 imaging objective that also focuses the laser beam. **b)** Schematics of single erythrocyte trapping by a laser hologram that reproduces its membrane contour. $\mathcal{E}(x,y)$ is the induced trapping potential. Red arrows indicate laser propagation, yellow arrows the optical forces. **c)** Image of an RBC in the equatorial plane without laser trapping. The membrane contour is retrieved as the high-contrast border between inner (black) and outer (white) circles. Scale bar is 2 μm.

### Single cell flickering landscape

To map the active flickering under physiological conditions we recorded fast-imaging videos of single free-standing RBCs lying on glass substrate [44]. To evaluate the flickering inhibition under holographic tweezing, a subsequent video was acquired while the static optical trapping potential $\mathcal{E}(r,\theta)$ was being applied. The reversibility of the process was proved by repeating this sequence more than once. To estimate the flickering activity we tracked the RBC membrane contour by exploiting detection schemes based on segmentation methods [45]. To remove spurious shifts, motions due to global cell translation or rotation were subtracted [46]. From the segmented membrane rim contour tracked as a function of time $R(\theta,t)$, we evaluated local radial fluctuations with respect to the average value, i.e., $\delta h(\theta,t) = R(\theta,t) - \langle R(\theta,t) \rangle_{\Delta T}$, where $\Delta T$ is the total recording time (see Methods and Supporting Information Fig. S2). For healthy human RBCs, the mean radius is $\langle R \rangle \approx 4$ μm [47], while the maximum local fluctuations are $|\delta h_{max}| \approx 0.2 - 0.3$ μm [48]. We define the time-averaged variance of the local membrane fluctuations as $\sigma^2_{\delta h}(\theta) \equiv \langle \delta h^2(\theta,t) \rangle_{\Delta T}$. In a first approximation, the active flickers explore a harmonic potential $U(\theta) = k(\theta)\sigma^2_{\delta h}(\theta)/2$, where $k(\theta)$ is the local membrane rigidity. By turning on the holographic optical tweezers, the RBC membrane is subject to the additional trapping potential $\mathcal{E} \propto I$ that increases the effective rigidity ($k_{trap} > k_{free}$). We compared the flickering activity of free-standing RBCs (tweezers off, $\mathcal{E} = 0$) and of the same cells subjected to a trapping potential deployed along the membrane contour (tweezers on, $\mathcal{E} > 0$).

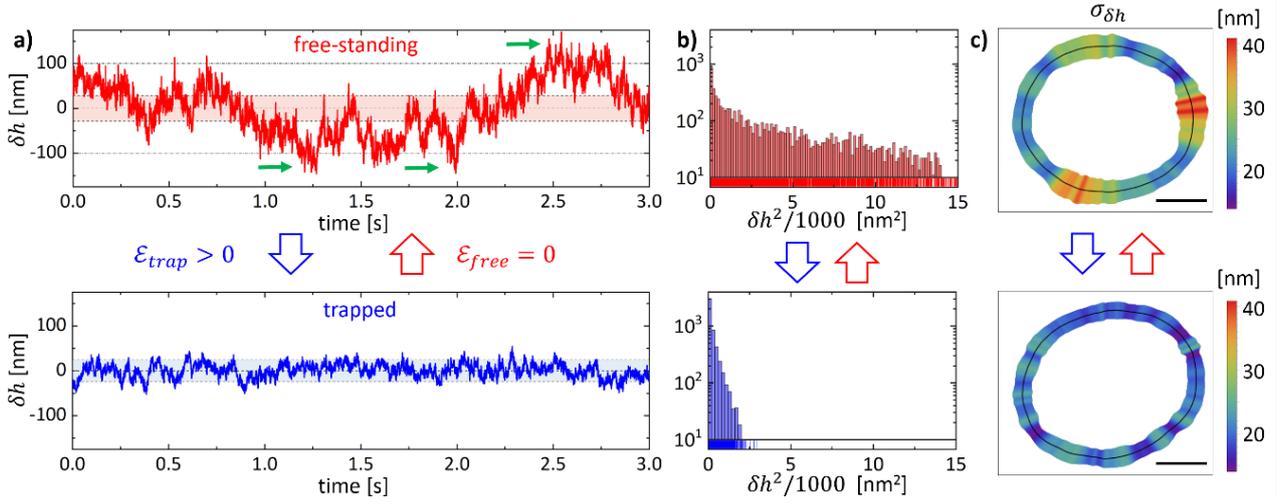

**Figure 2. Flickering mapping. a)** Time series of rim deformations $\delta h(t)$ tracked in a hot membrane spot of a free-standing RBC ($\mathcal{E} = 0$, red line), and in a cold spot of the same specimen under optically trapping ($\mathcal{E} > 0$, blue line). Coloured bands represent standard deviations $\sigma_{\delta h}$ of each time series. Green arrows highlight kicking events of amplitude $\Delta_0$ that can be larger than $3\sigma_{\delta h}$ (dashed lines). **b)** Distribution of $\delta h^2$ events for both cases. **c)** Map of the time averaged flickering activity along the cell contour estimated as $\sigma_{\delta h}(\theta)$ for the free-standing (top), and trapped RBC (bottom). Black line represents the mean position of the membrane. Scale bar is 2 μm.

Figure 2 shows a single RBC when free-standing (top panels) and under optical trapping (bottom panels). In Fig. 2a we report time series of membrane deformations $\delta h(\theta, t)$ tracked at representative rim emplacements. The flickering in free-standing RBCs exhibits large fluctuations ($|\delta h_{max}| > 3\sigma_{\delta h}$), whereas the trapped RBC membrane shows smaller deformations ($|\delta h_{max}| < \sigma_{\delta h}$). The free-standing cells display remarkable flickering volatility, characterized by transient pulses or localized membrane kicks, an activity fingerprint observed for a number of hot sites[44]. These transient kicking events involve large deformations $|\delta h| = \Delta_0 \approx 100$ nm sustained over relatively long times ($\tau_0 \leq 0.02$ s), as indicated in the upper panel of Fig. 2a (green arrows). In contrast, optical trapping not only decreases the fluctuation strength, but also inhibits the bursts of membrane activity, as illustrated in the lower panel of Fig. 2 a. Despite these differences, both flickering motions fluctuate around a zero- mean value $\langle \delta h \rangle_{\Delta T} \approx 0$, thereby demonstrating a stationary state of steady randomness. We highlight the difference in the fluctuation distribution of the two cases by reporting in Fig. 2b the number of events detected for each $\delta h^2$. The distribution of free-standing cells is broader than the one of the trapped RBCs, although they both decay nearly exponentially.

To evaluate landscapes of flickering activity along the cell contour, in Fig. 2c we show the membrane deformations as spatial maps of square-rooted variances $\sigma_{\delta h}(\theta)$, defined as standard flickering deviations [45,49]. These maps highlight that the deformation activity of the membrane depends on its trapping status and on local rim emplacement. For free-standing RBCs ($\mathcal{E} = 0$), variance maps reveal a heterogeneous mechanical activity featured by sparse hot spots, or membrane regions of small lateral size $a \leq 100$ nm with a high deformability $\delta h \geq 30$ nm (Fig. 2c top panel). These hot spots with enhanced mechanical activity can be considered as unitary membrane flickers, since their lateral size nearly resembles the dimensions of the triangular meshwork of the spectrin cytoskeleton where the protein that undergo phosphorylation are anchored in [50]. Moreover, the unitary kicker size agrees with an estimation based on previous observations (see Supporting Information). The stationary holographic trapping ($\mathcal{E}_{trap} > 0$) led to a significant reduction in flickering activity, as indicated by the smaller typical deformations $\sigma_{\delta h}^{(trap)} < \sigma_{\delta h}^{(free)}$, particularly in the warmer regions of the free-standing RBC. This condition also resulted in uniformly thermalized $\sigma_{\delta h}(\theta)$ maps, as reported in Fig. 2c bottom panel. In Supporting Information Fig. S3 we report examples of other RBC specimens that underwent passivated membrane deformations due to holographic tweezers trapping. We also observed that optical entrapment maintained $\sigma_{\delta h}(\theta)$ maps in a reversible manner (see Supporting Information Fig. S4).

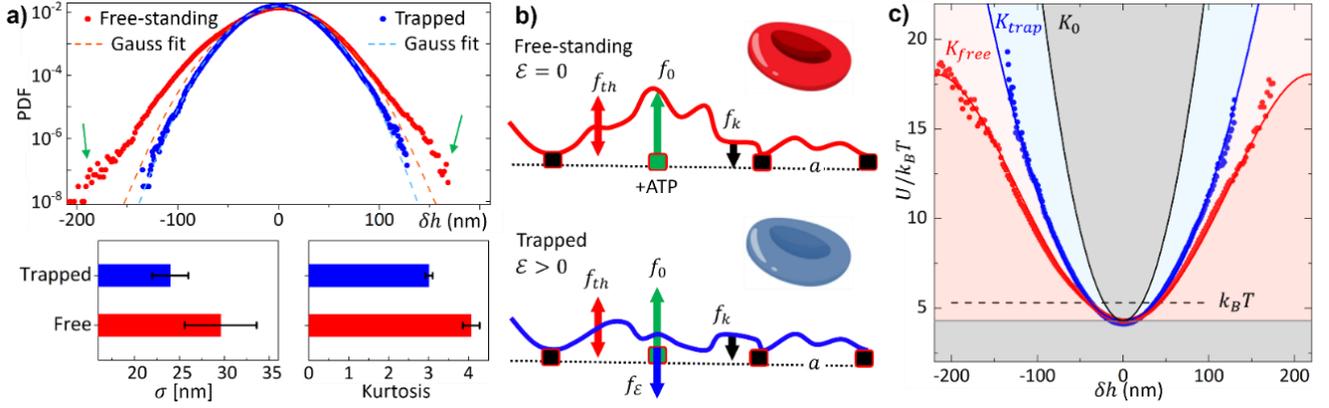

**Figure 3. Flickering landscapes. a)** Normalized probability density functions (PDFs) for free-standing ($\mathcal{E}_{free} = 0$, red dots) and trapped RBC population ($\mathcal{E}_{trap} > 0$, blue dots). Dashed lines are Gaussian fits performed considering only data in a short-fluctuation range [-50, 50] nm. In the free-standing population, long non-Gaussian tails correspond to active kickers, inhibited by optical tweezers. Lower panels show ensemble averaged flickering amplitudes ($\sigma$) and kurtosis, with error bars as standard deviations and p-value: $p = 0.018$, $p < 0.001$, respectively. **b)** Schematics of membrane deformations during flickering for free-standing (top panel) and trapped cell (bottom panel). The local force exerted by active kickers under ATP-release is $f_0 = k\Delta_0$; $f_{th}$ is the force induced by thermal fluctuations; $f_k$ is the restoring force due to membrane elasticity and $f_\mathcal{E}$ is the optical trapping force. **c)** Energy potential landscape in free-standing (red) and trapped (blue) RBCs, reproduced by Eq. (1) with $K_{free}$ and $K_{trap}$, respectively (straight lines). $K_0$ is the rigidity of passivated RBCs. Gray areas indicate exclusion regions where the flickering is absent. Dashed line corresponds to thermal driving energy $k_B T$.

To evaluate population differences, the probability density function (PDF) of each RBC ensemble was binned as number of flickering events occurred at a given deformation in any cell, equatorial emplacement and time, and then it was normalized to their integral. Figure 3 shows the PDFs of flickering deformations for the free and trapped ensembles, evaluated by accumulation of about $10^9$ microstates. Both PDFs are symmetrically distributed around $\langle \delta h \rangle = 0$. For free-standing RBCs, we found an average typical deformation $\sigma_{free} = (30 \pm 4)$ nm, in agreement with previous results [44,48,51]. The trapped population possesses a significantly lower deformation $\sigma_{trap} = (24 \pm 2)$ nm (see Methods for statistical significance). Nevertheless, trapped cells showed broader PDFs than those found as thermal fluctuations in passivated cells subject to ATP-inhibition $\sigma_{drug}^{(pass)} = 11$ nm, and also under tight membrane rigidization due to protein crosslinking $\sigma_{fixed}^{(pass)} = 8$ nm [44]. Therefore, we prove that holographic optical tweezers induce an effective trapping on RBC membrane while being able to maintain a flexible flickering status, provided that both populations remained biologically active. Dashed lines in Fig. 3a are tentative Gaussian fits of the PDFs considering only data in a small fluctuation range [-50, 50] nm, roughly corresponding to the thermal fluctuation domain [44]. The PDF of trapped RBCs sticks to the Gaussian fit also at larger $\delta h$, while for free-standing cells it deviates from the normal distribution, showing augmented tails. These non-Gaussian tails correspond to large membrane excursions represented by hots spots in $\sigma_{\delta h}(\theta)$ maps of free-standing RBCs in Fig. 2c. As expected for active kickers fuelled by glycolysis formed ATP [49,52], they can be assigned to outlier events $\delta h \gg \sigma_{free}$. These differences are statistically described by PDF's kurtosis $\mathcal{K} \equiv \langle \delta h^4 \rangle / \sigma_{\delta h}^4$, which is referenced to Gaussian distribution with $\mathcal{K}_0 = 3$ (see Fig. 3a, lower panels). For free-standing RBCs in glucose medium we found large kurtosis $\mathcal{K}_{free} = 4.1 \pm 0.2 > \mathcal{K}_0$ indicative of long tailed distribution, but also fingerprint of direct forces and anharmonic elasticity [53]. On the other hand, for trapped RBCs the flickering PDF is almost Gaussian since $\mathcal{K}_{trap} = 3.01 \pm 0.09 \approx \mathcal{K}_0$, hence indicating the inhibition of the active kickers and a recovery of harmonic elasticity, as it occurs for thermal fluctuations in rigidized RBCs [44].

## Flickering energy landscapes

In Fig. 3b) we model active flickering as a propulsion process in a flexible membrane anchored to a regular cytoskeleton defined in a lattice of period $a$, smaller than the kicking amplitude i.e., $a \leq \Delta_0$. We hypothesize the equatorial RBC-flickers as composed by $n = 2\pi\langle R \rangle/a$ flickering units either passive ($n_{pas}$), or active kickers ($n_{act}$), so that $n = n_{pas} + n_{act}$. The propelling motors become sparsely activated upon ATP-availability only in a small fraction of sites $\phi = n_{act}/n \ll 1$. The trade-off applied to a local flickering element gives a net force: $f_{flick} = \phi f_0 + f_{th} - f_k - f_\varepsilon$; akin stochastic propelling ($f_0$) and thermal ($f_{th}$) forces, by opposition to restoring elastic forces due to membrane rigidity ($f_k$) and optical trapping ($f_\varepsilon$), all of them resulting into the observed flickering action. In the harmonic approximation $f_{flick} = k\delta h$ (under local membrane hardness $k$) and $f_0 \approx k\Delta_0$, which determines the local kicking activity. By averaging over the membrane rim and population ensemble, the global flickering force emerges as $\langle f_{flick} \rangle = \langle k\delta h \rangle = K\sigma$, being $K = \langle k \rangle$ an effective global hardness. Since active flickering induces effective membrane softening, the global hardness is expected to be larger for the passivated ($K_0$) specimens with respect to living trapped ($K_{trap}$) and free-standing RBCs ($K_{free}$), thus following the relative succession $K_0 > K_{trap} > K_{free}$. To connect normalized PDFs with energy landscapes $U(\delta h)$, we assumed a Boltzmann-Gibbs statistics referred to the experimental temperature ($T = 37°C$): $PDF = \frac{1}{Z}\exp[-\langle U(\delta h) \rangle/k_B T]$ with the partition function $Z = 1$ due to normalization. Figure 3c shows the energy evaluated for free-standing and trapped ensembles. They are distributed upon the common ground energy $U_0 \approx 4\,k_B T$. For free-standing RBCs, we inferred an anharmonic potential, corresponding to large PDF kurtosis. To model the observed potential landscapes, we need to introduce active sources and nonlinearity into the harmonic potential. Since these membrane driving sources generate amplitude kicks only in a small subset of active nodes, we defined the fraction of active kickers using the dimensionless parameter $\phi$ ($0 < \phi \ll 1$). The kickers give rise to force of the order of $\langle f_0 \rangle \approx K\Delta_0$ and, therefore, to an active potential term $U_{act} = -\phi K\Delta_0|\delta h|$. Due to the symmetric distribution of the data reported in Fig. 3c, we argue that the active kickers also induce a nonlinear membrane softening that can be modelled by a nonlinear quartic potential term $U_{NL} = \phi\frac{1}{4}\beta\delta h^4$, where $\beta < 0$ is a softening parameter. Therefore, we fitted the measured potential energy of the living RBCs reported in Fig. 3c by the expression:

$$U = U_0 + \tfrac{1}{2}K\delta h^2 - \phi\left[K\Delta_0|\delta h| - \tfrac{1}{4}\beta\delta h^4\right] \tag{1}$$

The best fitting parameters that describe free-standing RBCs (considering $\Delta_0 = 100$ nm) are $K_{free} = (4.7 \pm 0.2)\,\mu\text{N m}^{-1}$, $\phi_{free} = (1.1 \pm 0.4)\cdot 10^{-2}$ and $\beta_{free} = (-8.2 \pm 0.5)\,10^9\,\text{Nm}^{-3}$. It follows that the kicking force is of the order of $f_0 \approx K_{free}\Delta_0 \approx 0.5$ pN. While for trapped RBCs we obtained an almost harmonic potential with $K_{trap} = (6.14 \pm 0.15)\,\mu\text{N m}^{-1}$, $\phi_{trap} = (3 \pm 1)\,10^{-3}$ and $\beta_{trap} = (-0.8 \pm 0.7)\cdot 10^9\,\text{Nm}^{-3}$. The trapping rigidity is larger with respect to free-standing RBCs but lower than the bare flexural rigidity of passivated RBCs $K_0 \approx 40\,\mu\text{N m}^{-1}$, as evaluated in[44] (see Supporting Information). Therefore, we deduce the membrane softness decreasing as the effective rigidity increases in the observed sequence $K_{free} < K_{trap} \ll K_0$. As a fingerprint of flickering passivation under trapping, the harmonic rigidization was evaluated nearly 25% (i.e., $(K_{trap} - K_{free})/K_{free} \approx 0.23$), that followed by the complete inhibition of the non-harmonic components ($\beta_{free} \gg \beta_{trap} \approx 0$, and $\phi_{free} \gg \phi_{trap} \approx 0$) revealed the practical hindering of the membrane activities as passivated by the holographic tweezers.

## Mean square deformation displacement

Further analysis to understand the RBC flickering stochastic process relies at a first instance on the mean squared displacements defined as $MSD \equiv \langle \delta h^2(\tau) \rangle_{\theta,N}$ (being $\tau$ the lag time between flickering events). Figure 4 shows ensemble averaged $MSD$s calculated for free-standing and trapped populations. A combined diffusion-confinement behaviour was observed in both cases across the characteristic viscoelastic time $\tau_C \approx 0.2$ s. At shorter lag times between flickering events ($\tau \ll \tau_C$), the ensemble-averaged flickering displacements can be reproduced by a generalized diffusion law $\langle \delta h^2(\tau) \rangle \approx 2D\tau^\alpha$, being the $D$ the effective diffusivity and $\alpha$ an apparent scaling exponent ($\alpha = 1$ for free diffusion, and $\alpha \leq 1$ for sub-diffusivity). As reported in the inset of Fig. 4, short-time sub-diffusivity was detected in both populations. The free-standing RBC flickers exhibit a limiting sub-diffusivity characterized at very short times by a power-law exponent $\alpha_{free} \approx 1/4$, whereas in this domain the trapped RBCs diffuse lesser ($D_{trap} \ll D_{free}$), under a higher exponent $\alpha_{trap} \approx 1/3 > \alpha_{free}$. No free diffusivity was observed in any temporal domain. Singularly observed in the free-standing flickers, we detected an active sub-diffusional crossover as a dynamic scaling renormalization undergone at $\tau_0 \approx 20$ ms (threshold for the faster kicking events). Arguably, the active sub-diffusivity is governed by the fastest transverse motions of the propelled kickers as they weakly relax under active membrane softness i.e., occurring at very short times ($\tau < \tau_0$). In fact, we observed the active kickers as uncorrelated permeation mode [54] ($\alpha_{free} \approx \alpha_{perm} \approx 1/4$ for $\tau < \tau_0$), which renormalize up to a mechanically correlated sub-diffusivity characterized by a slightly higher exponent ($\alpha_{free} > \alpha_{perm}$ for $\tau > \tau_0$). Henceforth, we will refer to this kicking decorrelation domain as producing short-time softness i.e., larger displacements than expected for fast sub-diffusive fluctuations. Conversely, the trapped RBCs exhibit a passivated sub-diffusivity that is closer to a flexural-like dynamics, highly correlated under membrane rigidness [55,56]. Such short-time flexural correlations are intermediate between conservative bending behavior in a pure elastic membrane leading to sub-diffusivity ($\alpha_{bend} \approx 2/3$), and that expected for hybrid curvature friction modes leading diffusivity leakages ($\alpha_{hyb} \approx 0$) [55,56]; otherwise stated we found that $\alpha_{hyb} \leq \alpha_{trap} \leq \alpha_{bend}$.

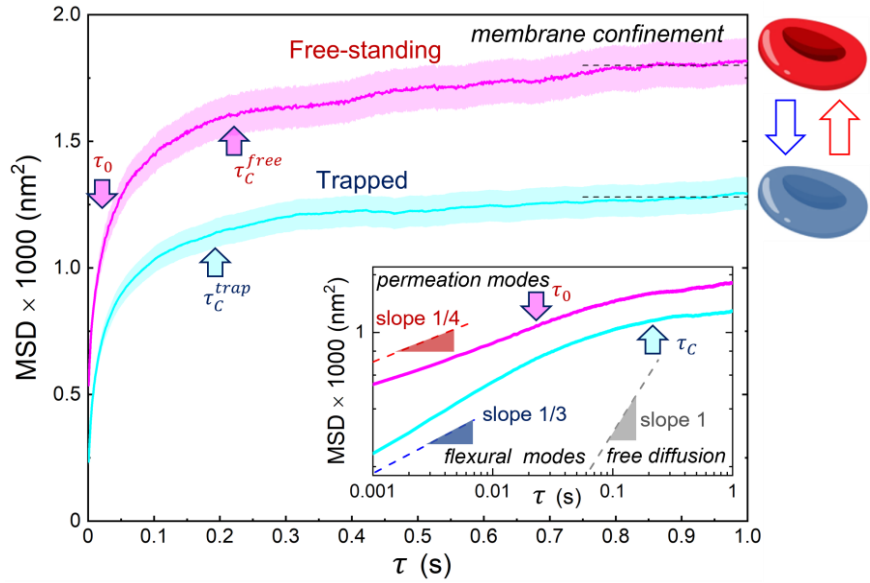

**Figure 4. Mean square displacement (MSD).** MSD averaged over membrane positions and RBC ensemble as a function of lag time $\tau$ for the free (purple line) and trapped (light blue line) RBCs along with corresponding 90% variability bands. The microscopic time $\tau_0 \approx 20$ ms (threshold for instantaneous flickering events) and $\tau_C \approx 0.2$ s (viscoelastic correlation time) are highlighted by arrows. The inset shows the log-log plot of MSD compared with characteristic slope decay for free diffusion (gray line, $\alpha = 1$), membrane flexural modes (blue line, $\alpha = 1/3$) and membrane permeation modes (red line, $\alpha = 1/4$).

Rigidity-confinement correlations of flickering displacements occur at longer lag times than the characteristic viscoelastic time ($\tau > \tau_C$). In this case, the flickering displacements asymptotically reach a saturation plateau due to membrane confinement under effective flexural rigidity as $\langle \delta h^2 \rangle_{\tau \to \infty} \approx k_B T / 2K$, where $K$ is the effective membrane tension acting as a flexural modulus [55,57]. For the free-standing RBCs, we found $\langle \delta h^2 \rangle_{\tau \to \infty}^{free} = (1800 \pm 120)$ nm², which implies a softened flexural stiffness $K_{free} = (4.8 \pm 0.3)\mu$N m$^{-1}$, in agreement with the effective spring constant obtained from the energy landscape of Fig. 3. For the trapped RBCs, however $\langle \delta h^2 \rangle_{\tau \to \infty}^{trap} = (1280 \pm 90)$ nm², so that $K_{trap} = (6.7 \pm 0.5)\mu$N m$^{-1} > K_{free}$, in agreement with the model reported in Fig. 3. The characteristic softening activity formerly observed in free-standing RBCs, here appears inhibited as reveled by the increase of the confinement plateau. Provided the holographic tweezer is switched on, the trapping stiffness hinders the most active kickers, so that the active softening due to membrane kicking drops, whereas the terminal confinement rises. When the tweezer is turned off, however, the membrane becomes once again free to flicker under active softness fuelled by kicking propulsion, so that the lower confinement is reversibly retrieved as followed by the recovery of the characteristic kicking dynamics at short times.

**Power spectral density**

To compare the active RBC flickering with the passive Brownian noise, we studied the spectral components of the fluctuation series. The power spectral density ($PSD$) was obtained by Fourier transforming the flickering signals: $PSD(\omega, \theta) = (2\pi)^{-1} \| \int_0^\infty \delta h(t, \theta) e^{-i\omega t} \, dt \|^2$. Figure 5 shows the average $PSD$ for free-standing and trapped RBC populations. In both spectra, we found a terminal domain where a collective activity plummets for $\omega > \omega_A \approx \tau_c^{-1} \approx 4 \, Hz$ (viscoelastic threshold), and a crossover frequency ($\omega_C \approx \tau_0^{-1} \approx 35 \, Hz$). The collective power of the free-standing RBCs occurs in the low frequency regime ($\omega < \omega_A$), although biological activity also reemerges in the high-frequency spectral tails ($\omega > \omega_C$), which evidenced the rapid propulsion of the individual kicking events. We found that active free-standing RBCs generate stronger kicking power at higher frequencies ($\omega > \omega_C$). In this frequency range we detected a pink noise spectrum typical for correlated noise corresponding to membrane kicking stochasticity ($PSD \sim \omega^{-1}$) [58]. Since in the intermediate range ($\omega_A < \omega < \omega_C$), $PSD$ appears independent on the trapping status, we envisaged a rheologically passive regime [44,51,59]. This corresponds to the thermal excitement of mechanical modes in a passive rigid continuous that dissipate energy via bulk friction as exhibiting $PSD \sim \omega^{-5/3}$ [51]. Here, we invoked the Milner-Safran (MS) theory for the curvature fluctuations of passively deformable vesicles [60,61]. Oppositely, we observe the collective activity of the free-standing RBCs generating the highest power at the lowest frequencies ($\omega < \omega_A$). The collective metabolic activity articulated under long-range elastic correlations in the RBC-membrane can be included as an additive Lorentzian contribution in this low frequency domain ($\omega < \omega_A \approx \tau_c^{-1}$), which corresponds to the relaxation terminus where the membrane behaves essentially frictionless [44,51]. Therefore, the shape fluctuations of the studied RBCs are expressed as a superposition of effective MS- modes (defined by spherical harmonics $l = 2, 3, 4 \ldots$), with an additional kicking mode describing collective membrane activity [62]:

$$PSD(\omega) = \frac{\langle R \rangle^2}{4\pi^2} \left[ \sum_{l \geq 2} \frac{2l+1}{2\pi} \cdot \frac{\omega_l}{\omega^2 + \omega_l^2} A_l^{MS} + \frac{\omega_A}{\omega^2 + \omega_A^2} A^{act} \right] \quad (2)$$

where $\langle R \rangle$ is the average membrane radius; the MS-amplitudes are $A_l^{MS} = k_B T [\kappa(l+2)(l-1)l(l+1) + \sigma R^2(l+2)(l+1)]^{-1}$, defined in terms of the bending modulus $\kappa$, and the membrane tension $\sigma$; the MS-frequencies are $\omega_l = \kappa(l+2)(l-1)l(l+1) + \sigma R^2(l+2)(l+1)[\eta R^3 \Lambda(l)]^{-1}$, being $\eta$ the bulk viscosity, and $\Lambda(l) = (2l+1)(2l^2 + 2l - 1)[l(l+1)]^{-1}$. The collective activity is described by a dimensionless parameter accounting for the global kicking amplitude $A^{act}$, appeared at a metabolic turnover frequency $\omega_A$ [62,63].

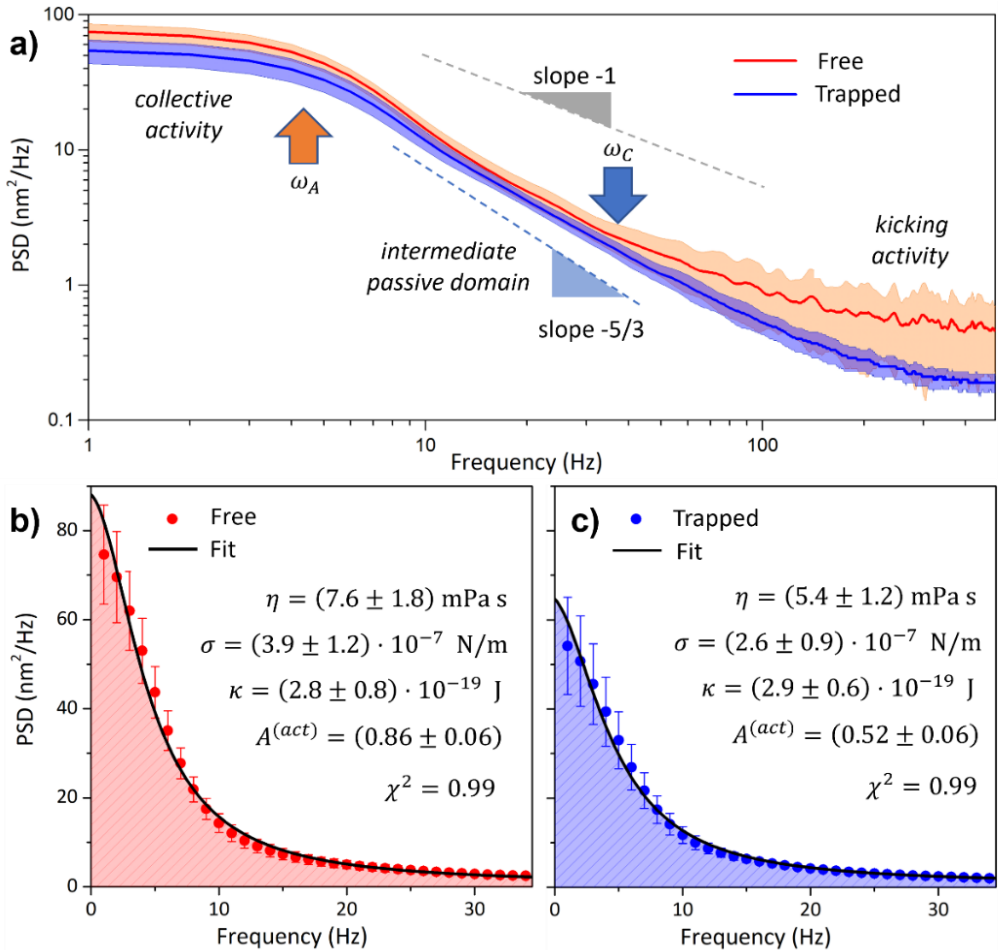

**Figure 5. Power spectral density (PSD)** of flickering obtained by Fourier transforming the time series $\delta h(t)$. **a)** Ensemble-averaged PSDs over RBC free-standing (red line) and trapped population (blue line). Shaded areas represent the 90% variability intervals. Dashed lines decay as $\omega^{-1}$ (pink noise spectrum), or as $\omega^{-5/3}$ (membrane Brownian spectrum). Arrows highlight the active frequency $\omega_A$, and the rheological crossover frequency frequency $\omega_C$. **b)-c)** Averaged PSD of the membrane fluctuations in the low frequency regime for the free-standing (red dots). and trapped RBCs (blue dots). Error bars correspond to the standard deviation at each frequency. Black lines are the fitting by using Eq. (2), with the dashed region corresponding to the integrated power. Fitting parameters are reported for each case.

On the one hand, the discrete sum of normal modes in the MS- spectrum [60] describes the relaxation of the RBC's shape changes governed by flexural curvature modes, through the effective membrane tension and bending modulus at intermediate frequencies ($\omega_A < \omega < \omega_C \approx 35$ Hz). On the other hand, the active term determines a Lorentzian mode of corner frequency $\omega_A \ll \omega_l$, and amplitude $A^{act}$, which is collectively driven by a relatively small number of active flickering units ($n_{act} \ll n$) [64]. In Fig. 5 b-c we focus on the frequency domain where collective activity is dominating and the PSDs data were fitted by Eq. (2) by imposing $\langle R \rangle = 4.0$ μm, $T = 37°C$, and $\omega_A = 4.3$ Hz (see Fig. 5b-c). The best fitting values for the passive mechano-structural parameters are reported in the insets together with associated uncertainties ($\sigma$, $\kappa$ and $\eta$). It is worthy to stress that the passive parameters are consistent between free-standing and trapped populations ($\eta \approx 6$ mPa s, $\kappa \approx 70\ k_B T$, $\sigma \approx 0.3$ μN m$^{-1}$), and with previous analyses under similar metabolic conditions [44,51]. The amplitude of the active term is also in agreement with previous studies [45]. For the free-standing RBCs, the dimensionless amplitude was found to be $A_{free}^{act} = 0.86 \pm 0.06$. The collective activity is meaningfully decreased down to $A_{trap}^{act} = 0.52 \pm 0.06$ in the trapped cells. Moreover, we obtained an integrated power of 590 nm$^2$ for the free-standing population and of 450 nm$^2$ for the trapped RBCs thus indicating that the dynamic activity of the living cells was reduced by ca. 25% under optical trapping. As aforementioned, a similar conclusion can be achieved from the analysis of the static PDFs.

A dynamically mappable characteristic of membrane activity that can be extracted from the flickering series is the effective diffusivity obtained by fitting the local short-time displacements to the Einstein relationship in one-dimension. For $\tau < \tau_0$, this is $MSD(\tau, \theta) = 2D(\theta)\tau$. Figure 6a-b show the $\sigma_{\delta h}(\theta)$, and diffusivity, $D(\theta)$, maps obtained along the membrane rim for a representative RBC in the free-standing status (left panels), and the corresponding trapped case (right panels). Analogously to $\sigma_{\delta h}(\theta)$, the diffusivity maps of free-standing RBCs exhibit a heterogeneous distribution, with hot spots of localized activity characterized by fast membrane displacements, which coexist with colder regions corresponding to lower diffusivities. Additional $D(\theta)$ maps for different RBCs are reported in Supporting Information Fig. S5. To evaluate population averaged behaviours, Fig. 6c shows densities of states into heat maps of membrane activity recorded as the kinetic diffusivity ($D$) versus the elastic deformability ($\sigma^2_{\delta h}$). This allows us to highlight differences between the two flickering ensembles, since free-standing RBCs (left panel) show a more widespread map due to strong diffusion with $D$ extending beyond $D > 15$ nm$^2$ms$^{-1}$ and large displacement variance $\sigma^2_{\delta h} > 800$ nm$^2$. While trapped RBCs (right panel) show more localized maps with shorter displacements for the vast majority confined in the range $D < 10$ nm$^2$ms$^{-1}$ and $\sigma^2_{\delta h} < 800$ nm$^2$. Moreover, to quantify the distribution spreading, the degree of linear correlation between the variables $D$ and $\sigma^2_{\delta h}$ can be estimated by the Pearson coefficient $P(D: \sigma^2_{\delta h})$ [65]. A higher correlation between kinetic ($D$) and potential ($\sigma^2_{\delta h}$) parameters is found in the trapped population ($P_{trap} = 0.74$), with respect to the free-standing RBCs ($P_{free} = 0.49$), hence indicating confined behaviour and reduced activity in the trapped cells. Whenever the living RBCs are maintained free-standing, the higher diffusivities corresponding to the more active motions are observed uncorrelated from the passive rigidity of the membrane.

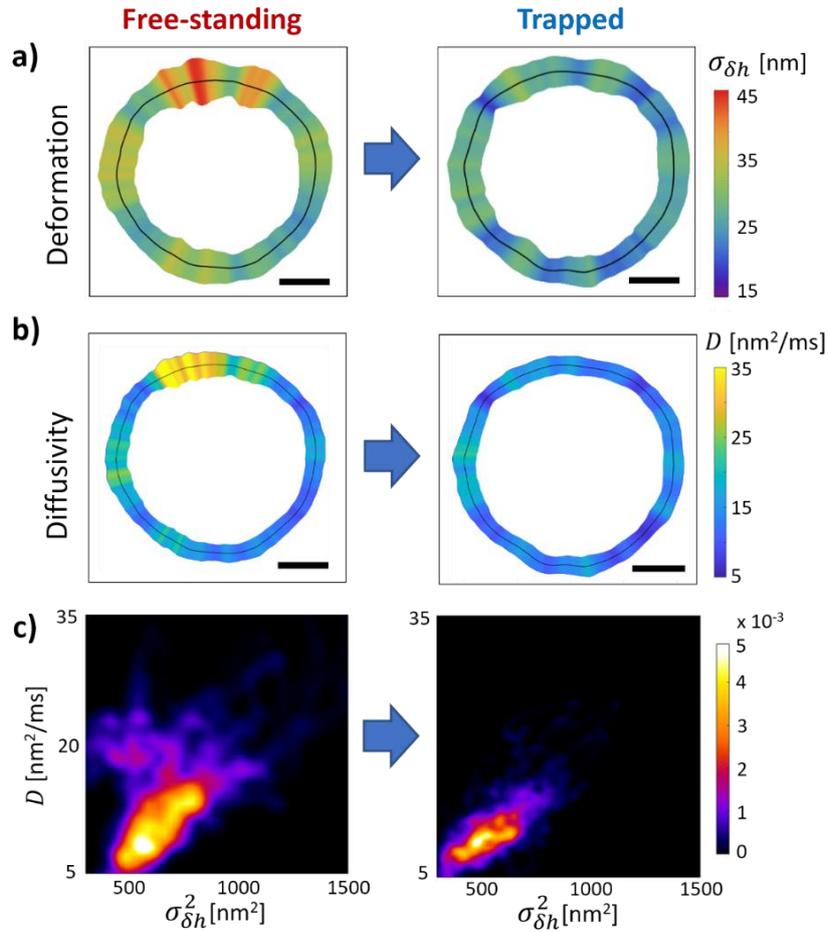

**Figure 6. Diffusivity and deformability maps of RBC flickering. a)- b)** Maps of deformation variance and membrane effective diffusivity $D$ of the same specimen, respectively, for free-standing RBC (left panels) and trapped RBC (right panels). Black lines represent the mean position of the membrane. Scale bar is 2 μm. **c)** Probability distribution of membrane spot observed in the coordinates $(D, \sigma^2_{\delta h})$ for free-standing population (left panel) and trapped RBCs (right panel).

**Discussion**

We developed a holographic optical tweezers to induce landscapes of trapping potential along the equatorial RBC membrane rim. The stationary optical potentials were designed *ad hoc* under a feedback control of the observed cell. The system operates at minimal invasiveness, without introducing force transducer microspheres, that may request chemically complex membrane attachments and invasive procedures. Moreover, the trapping process showed negligible phototoxicity, without strongly altering cell size and shape, as proven by the reversibility of the observed flickering deformations. Although our holographic tweezers are not yet equipped with a direct force measurement, membrane flicker mapping allows estimating induced strengths of trapping potential by evaluating deformation PDF distributions of free-standing, trapped and passivated cells. The flickering PDF-landscapes underlie causative forces restored by membrane rigidity, that can be driven by active process (free-standing cells) or passive Brownian-like motion (passivated cells). The $\sigma_{\delta h}$ and diffusivity maps highlight the differences of flickering in different regions of single RBCs, being statistical significative when compared the ensemble average free-standing cells with the same trapped specimens. When holographic tweezers were turned on, we detected an increased membrane stiffening ($K_{trap} > K_{free}$), hence indicating active kicking becoming hindered. We estimated the stiffness induced by optical trapping as $\Delta K = K_{trap} - K_{free} \approx 1.4$ μNm$^{-1}$ per cell, being of the same order of magnitude than the erythroid membrane rigidities previously evaluated by methods of flickering spectroscopy [17,44,51,59,66].

Both, the flickering displacements observed in time domain (from MSD in Fig. 4), and the dissipated power evaluated in frequency domain (form PSD spectra in Fig. 5) show signatures of activity in the free-standing RBC population. In the high-frequency range ($\omega > \omega_c$), we observed high-power dissipation at high rates. This was also evidenced in MSDs as an active softening led to larger flickering than expected at very short times. The associated active modes presented a higher deformation strength in the free-standing RBCs, which disappeared in the trapped ones. Therefore, we were able to address the activity of individual kickers. At longer lagging between flickering events ($\tau > \tau_c$), which correspond to lower flickering frequencies ($\omega < \omega_A$), we detected a collective mechanical activity associated with the sparse kicking events, that are averaged over long observational times and are highlighted by hot regions in the spatial maps of membrane deformability. This collective activity corresponds to the equilibrium breakdown that was reported as a violation of the fluctuation-dissipation relation, which is a demonstration of the non-equilibrium nature of flickering in living RBCs [48]. However, in the intermediate-frequency range ($\omega_A < \omega < \omega_c$), both dissipation spectra decay as $\omega^{-5/3}$, corresponding to a Brownian response due to thermal motion. Regarding the effects of holographic tweezing, although a notable reduction in flickering activity of RBCs was observed during the trapping of membrane kickers along the equatorial rim, the trapped RBCs do not display noticeable modifications in cellular morphology (size and shape). Additionally, there are no substantial changes in the measured mechanical properties, including membrane tension, rigidity, and viscosity. This non-invasive optical manipulation was reversible, with all physical and biological degrees of freedom effectively restored once the tweezers are deactivated.

By assuming that the collective activity is dictated by the membrane kickers, then $A^{act} \sim n_{act}$ [64]. Then, from the $PSD$ analysis (Fig.5) the ratio of decreased activity due to optical trapping can be defined as $\theta_{trap} = \left(A^{act}_{free} - A^{act}_{trap}\right)/A^{act}_{free} \approx 0.4$, which indicates that a majority of kicking elements (nearly 60%) were passivated under trapping. This estimation corroborates our previous estimation $(\phi_{free} - \phi_{trap})/\phi_{free} \approx 0.7$. The decrease of the collective activity is compatible with the hindering of the rapid membrane kicking motions due to holographic optical tweezers, as evidenced both in the MSDs and in the PSDs.

In free-standing RBCs active kickers that dissipate kinetic energy appear as biologically hot regions heterogeneously distributed along the membrane, in agreement with previous finding indicating asymmetries in the cytoskeletal complex [67]. In contrast, optically trapped RBCs exhibit reduced kinetic energy dissipation and can be considered biologically cold. The distinction between active and trapped RBCs is also reflected by softer membrane found in free-standing cells, which leads to elastic decorrelations between $D$ and $\sigma_{\delta h}^2$, as reported in Fig. 6. Therefore, mechanically softened, and thermodynamically hot RBCs do behave out-of-equilibrium as far as they are the metabolically active. Conversely, physical correlations dominate in thermodynamically colder RBCs under trapping, indicating a mechanical balance between elastic and dissipative energy. Arguably, the biological inhibition give rise a kind of quiescent state in the trapped RBCs.

In conclusion, by developing a custom-designed holographic optical tweezers, we obtained a minimally invasive, non-phototoxic, and biocompatible tool capable of precisely manipulating individual erythrocytes by directly applying mechanical forces to the plasma membrane. Our approach enables the application of spatially extended and dynamically adaptable patterns of optical potentials. In a proof-of-principle demonstration, we manipulated the flickering activity along the membrane of RBCs, while preserving their structural characteristics. We found active kickers heterogeneously distributed along the plasma membrane of living RBCs and we hindered their flickering by applying an external force field along the membrane rim. Furthermore, a MATLAB-based algorithm was customized to enable nearly real-time manipulation of the membrane contour, establishing a foundation for future extensions to more complex geometries and diverse spatiotemporal force field configurations. Our findings validate optical manipulation as valuable tools for locally controlling living cell membranes. Our approach holds potential in shaping innovative strategies for achieving precise mechanical regulation in living cells. For instance, it could contribute to control force fields during pivotal cellular processes, such as cell division or nuclear cytokinetic reorganization, thus exerting a significant influence on cellular mechanics in various biological contexts.

**Materials and Methods**
**Microscopy setup.** The illumination source of white visible light is a LED (WFA1010, by Thorlabs) whose beam is collimated by lens $L_1$ (with focal length $f_1$=10 cm) and then focused on the sample plane by a condenser objective Ob1 (5X, NA=0.25, by Nikon). The sample is mounted on a two-axis (x,y) translation stage (by Newport) for precise alignment at the sub-micron scale. The collection objective Ob2 is a 100 X achromat oil-immersion objective with NA=1.45 (by Nikon). It is placed on a single axis (z) translational stage (by Thorlabs) with 10 nm resolution. The collected light is transmitted through a dichroic mirror (D, short-pass at 650 nm, which transmits visible light and reflects near-IR light) and finally focused by the tube lens $L_2$ (with focal length $f_2$=20 cm) on a CCD camera or on a high-speed CMOS camera (PROMON U750, by AOS Technologies) by switching a movable mirror $M_2$. To filter out the residual laser reflection on the sample from the image, an additional low-pass filter ($\lambda < 700$ nm) is placed in front of both camaras. The experiments were performed acquiring images at 1 KHz repetition rate and a total of 5000 frames for each sample. Overall, we estimate the shortest flickering time in the Nyquist limit $t_0 \approx 2$ ms.

**Holographic optical tweezers.** A continuous-wave laser diode emitting at 785 nm (ONDAX, by COHERENT), operating at a stable output power of 150 mW was used as light source. We control the laser power by acting on a rotating half-wave plate to change the polarization of the beam, in conjunction with a linear polarizer. This allows us to fix the polarization (with electric field parallel to the larger dimension of the SLM screen) and the power delivered to the sample plane for achieving optical trapping at 10 mW. A telescope with magnification 8 X (made by lenses $L_3$ and $L_4$, with $f_4$=2.5 cm and $f_3$=20 cm) collimates and increases the size of the laser beam waist. An iris $P_1$ selects the central part of the beam, in order that a homogeneous beam with and constant wavefront impinges on the central area of the SLM (LCoS Pluto 4.0, by HOLOEYE) that works in reflection. Each SLM's pixel (1980 x1024) is set to induce an independent phase shift between 0 and $2\pi$, controlled by applied voltage defined via computer. To draw on the sample plane the computer-generated holograms, the SLM was placed in the Fourier plane of the sample. The SLM response is programmed by MATLAB software to induce a phase mask that sculptures the laser phase to accomplish two tasks: *i)* draw an intensity pattern that resemble the membrane of the RBC observed by the camara; *ii)* act as a virtual lens with focal length $f_V$=38 cm. This virtual lens along with the lens $L_3$ ($f_3$=30 cm, placed at 68 cm to the SLM and 30 cm to the objective back aperture) form a relay optics that images the beam reflected at the SLM plane to the back-focal plane of Ob2. The beam size overfills the objective back aperture and generates a strong optical trap in the sample plane. In the optical path the laser is reflected by the dichroic mirror before entering the objective.

**Hologram generation.** The hologram focused on the sample plane reproduces the membrane of the observed RBC. From the image acquired by the camera in real time, we evaluated the profile of the investigated RBC membrane by generating a 1920x1080 binary image of the cell rim. To achieve this goal, we developed a MATLAB based software, whose relevant input/outputs are reported in Supporting Information Fig. S1. Then, by using the binary image, we calculated its phase distribution by employing the Gerchberg-Saxton algorithm [68]. This phase-retrieval algorithm is based on Fourier-transform loops that converge to display an 8-bit phase distribution image with the same dimension of the SLM screen. The phase distribution was then displayed on the SML screen to modify the incoming phase-homogeneous laser beam. Moreover, the virtual lens is programmed in the SLM by adding to the phase mask calculated by Gerchberg-Saxton algorihm the phase distribution of a lens with focal length 38 cm and finally unwrapping the total phase mask in the range $[0, 2\pi]$. Finally, the laser beam reflected by the SLM is focused on the sample plane and generates the intensity pattern resembling the observed RBC membrane.

**Sample preparation.** RBCs were extracted from healthy donors by venipuncture employing the finger-prick method. 20 µl of blood were resuspended in 450 µl phosphate saline buffer (130 mM NaCl, 20 mM $Na_3PO_4$, 10 mM glucose, and 1 mg $mL^{-1}$ bovine serum albumin, pH 7.4) at 37 C and denoted as PBS+. The erythrocyte concentrate was obtained after 3 centrifugations (each cycle of 10 min at 5000 *g*). The supernatant was discarded, and the pellet rinsed in PBS+ (450 µl) after the first and the second centrifugation, after the third the pellet is resuspended (1:15) in PBS+. From this suspension, a 1:20 dilution was stored at 37°C. Samples were mounted on microscope glass slides with a square-shaped well in the middle of the slide made of double-side adhesive tape. Aliquots of 20 µl from the 1:15 suspension were poured into the well and a thin cover-glass was placed on top and thus sealed by the adhesive tape. Therefore, the sample chamber is made of a microscope slide and coverslip that are connected through a double-sided sticking thin film, with total volume equal to 1 cm x 1 cm x 50 µm. To avoid RBCs adsorption to the glass surfaces of the chamber, both the well and the cover-glass were coated with bovine serum albumin (BSA, 40 mg $mL^{-1}$).

**RBC trapping protocol.** We consider the living RBC horizontally lying by gravity on the microscopy slide. For free-standing flickers, we recorded videos in which RBCs undergo flickering motions in healthy physiological conditions. Once the desired phase mask of the optical intensity of the observed RBC membrane contour was digitally encoded in the SLM, we activated the distributed holographic optical tweezers by turning on the laser and delivering a power of 10 mW on the sample plane. Laser power above 10 mW proved to be sufficient to induce a flickering inhibition. For the trapped cell we recorded a video with the same parameters used previously. To assure that the laser trapping does not induce irreversible changes and prove reproducibility, for every sample we performed the following sequence of experiments: 1) video of free RBC flickering fluctuations; 2) video of RBC flickering developed under optical trapping; 3) video of free RBC.

**Flickering analysis.** Each frame of the video acquired by the fast camera is analysed by MATHEMATICA based software to evaluate the position of the circular RBC membrane as a function of time. The rim halos that move at the RBC membrane allow to estimate the radial positions of the contours by interpolation at subpixel resolution with an accuracy of few nm, based on a segmentation algorithm developed in [45]. We recorded the time series of the local membrane fluctuations for 2048 segmented rim elements, corresponding to an angular step $\Delta\theta \approx 3.1 \cdot 10^{-3}$ rad, and to an arc length of $12\ nm$ that is smaller than the estimated size of the flickering unit. An example of membrane profile estimation is reported in Supporting Information Fig. S2. By tracking the membrane position over time, we evaluate the fluctuations $\delta h(t, \theta)$ and correct the displacement by subtracting the global drift that the RBC may experience during the recording due to global translation and rotation as in [46]. We evaluate the best alignment between consecutive frames, i.e., the rigid transformation that conserves the distances among the traced membrane elements. The positions of the centre of mass were considered as the drifting trajectory to be subtracted to the coordinates of any displacement.

**Statistical analysis.** Regarding the values of $\sigma$ and kurtosis estimated in Fig. 3, we performed the t-test to assess the statistical difference between the two groups of free and trapped RBCs in $N = 12$ specimens. For a level of significance lower than 5% for the null hypothesis ($p < 0.05$), we observed $p = 0.018$ for $\sigma$ and $p \approx 10^{-6}$ for kurtosis. When fitting procedure was performed the goodness of the nonlinear fits was proven by the high likelihood $\chi^2$-estimator.

**Data availability:** The data that support the findings of this study are available from the corresponding author upon reasonable request.

**Scripts.** The scripts and functions used for functioning the SLM, acquiring and analysing the data were built on MATLAB (R2020b) and MATHEMATICA (12.1.0).

**Acknowledgments:** The authors acknowledge Comunidad de Madrid for funding this research under grants Y2018/BIO-5207 and S2018/NMT-4389, and Spanish Ministry of Science and Innovation (MICINN) – Agencia Española de Investigación AEI under grant PID2019-108391RB-100 and TED2021-132296B-C52. This study was also funded by a REACT-EU grant from the Comunidad de Madrid to the ANTICIPA project of Complutense University of Madrid. The funders had no role in study design, data collection and analysis, preparation of the manuscript or decision to publish.

**Author Contributions:** N.C. conceived the experiments, mounting the optical setup, and analyzing the data, with F.M. designing the research. N.C. and M.G.V. conducted the experiments, supported by M.C. and N.H.O. in sample preparation, with F.M. designing both conceptualizations. The manuscript was written by N.C. and F.M.

**Competing Interest Statement:** The authors declare no competing interest.

# Supporting Information for
Optical control of spatially localized red blood cell activity
by holographic tweezing

Niccolò Caselli[1,2], Mario García-Verdugo[1], Macarena Calero[1,2], Natalia Hernando-Ospina[1,2], José A. Santiago[3], Diego Herráez-Aguilar[4] and Francisco Monroy[1,2]

[1] *Departamento de Química Física, Universidad Complutense de Madrid, Ciudad Universitaria s/n, 28040 Madrid, Spain.*
[2] *Translational Biophysics, Instituto de Investigación Sanitaria Hospital Doce de Octubre, 28041 Madrid, Spain.*
[3] *Departamento de Matemáticas Aplicadas y Sistemas, Universidad Autónoma Metropolitana Cuajimalpa, Vasco de Quiroga 4871, 05348 Ciudad de México, México.*
[4] *Instituto de Investigaciones Biosanitarias, Universidad Francisco de Vitoria, Ctra. Pozuelo-Majadahonda, Pozuelo de Alarcón, Madrid, Spain.*

Corresponding author: Niccolò Caselli. Email: ncaselli@ucm.es

**This PDF file includes:**
  Supporting text
  Figures S1 to S5
  SI References

**Supporting Information Text**

**Unitary flickering element**

In order to estimate the unitary flickering size ($a$), we modelled the flickering deformations in terms of an effective membrane hardness, by invoking the Helfrich's theory of membrane mechanics as grounded on the density of flexural stiffness $K$ [1]. The minimal level of flickering variance ($\sigma_0^2$) can be retrieved from the fluctuation-dissipation balance between thermal driving and the restoring elasticity of the rigid membrane found in previous experiments with fully passivated RBCs subject to ATP-inhibition $\sigma_{drug}^{(pass)} = 11$ nm, or under membrane rigidization by protein crosslinking[2] $\sigma_{fixed}^{(pass)} = 8$ nm. Therefore, we can define $\sigma_0 \sim 10$ nm and $\sigma_0^2 \sim 100$ nm$^2$.

Using the equipartition energy theorem $K_0 \sigma_0^2/2 = k_B T/2$, therefore, the bare flexural hardness is $K_0 = k_B T/\sigma_0^2 \approx 40$ μN m$^{-1}$ at the operating temperature $T = 37°C$.

Moreover, $K_0$ can be related to the unitary flickering size $a$ by $K_0 \equiv \kappa_0/a^2$. Consequently, being the rigidity modulus corresponding to the bare bending rigidity per flicker unit $\kappa_0 \approx 70\ k_B T$ [2,3]. This allows us to estimate $a \approx \sqrt{\kappa_0/K_0} \approx 80$ nm.

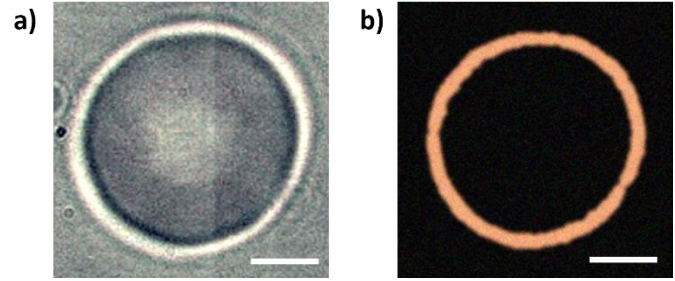

**Fig. S1.** Red blood cell (RBC) membrane detection by MATLAB and holographic trap generation. a) Optical image of a healthy RBC in PBS+ solution deposited on glass substrate without laser trapping. b) Membrane contour generated from the MATLAB detection algorithm. This image was used as input for calculating the phase mask to be displayed on the spatial light modulator (SLM). Scale bar is 2 μm.

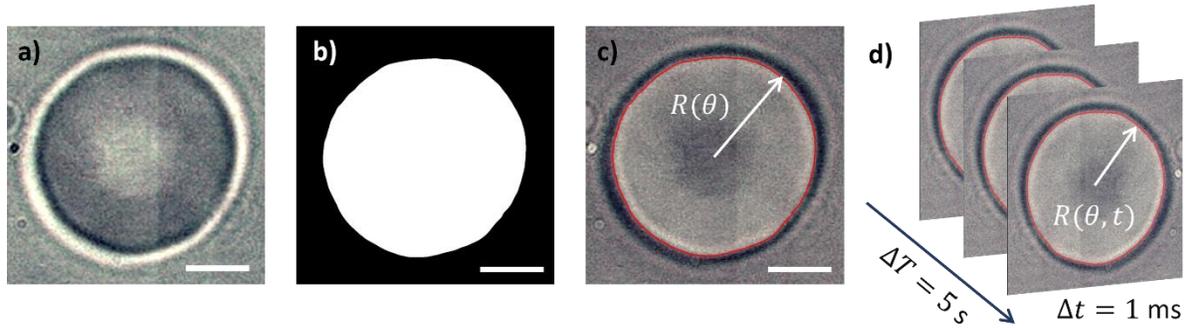

**Fig. S2.** RBC detection by MATHEMATICA and tracking of membrane fluctuation. a) Optical image of the RBC in PBS+ solution without laser trapping acquired with exposure time of 1 ms. b)-c) Output of the MATHEMATICA membrane detection algorithm that show a binarize image and a red circle representing the contour of the evaluated membrane, respectively. Scale bar is 2 μm. Membrane detection allows for evaluating the radius of the membrane $R(\theta)$ in a single image. d) By tracking the membrane position in successive frames over time $R(\theta, t)$ was retrieved, with a temporal resolution of 1 ms.

Then the instantaneous fluctuations were defined as $\boldsymbol{\delta h(\theta, t) = R(\theta, t) - \langle R(\theta, t) \rangle_{\Delta T}}$.

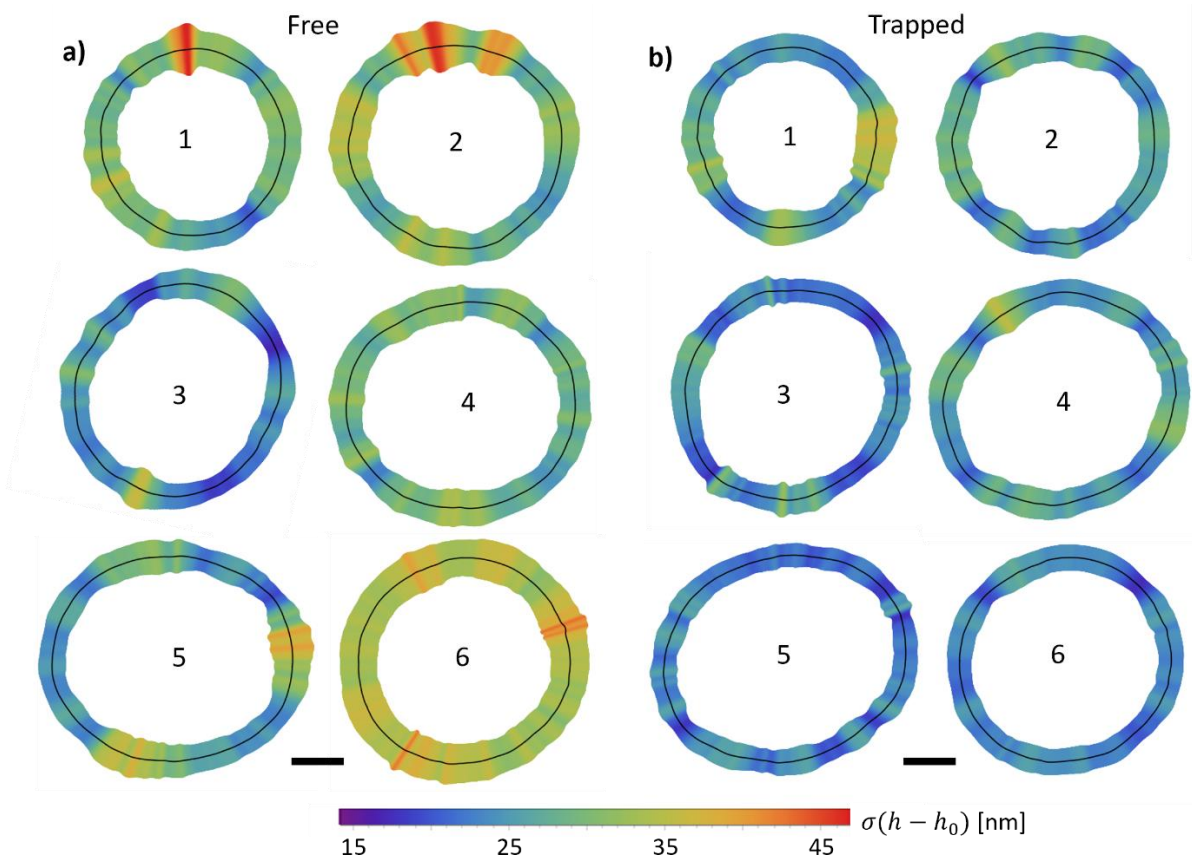

**Fig. S3.** Flickering fluctuation maps of single RBCs. a) RBCs free to oscillate are labelled from 1 to 6. The coloured circular maps represent for each angle the standard deviation $\sigma$ of the fluctuation $\boldsymbol{\delta h(t,\theta)}$. Black lines represent the mean position of the membrane. Scale bar is 2 μm. b) The same RBCs reported in a) are analysed when subject to a trapping potential distributed along the membrane circle with 10 mW at the sample plane.

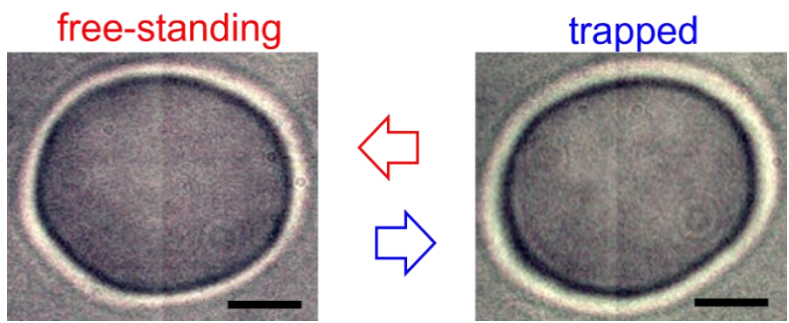

**Fig. S4.** Instantaneous phase-contrast image of a healthy RBC when free standing on a glass substrate (left) and when trapped by holographic optical tweezers (right). Scale bar is 2 μm.

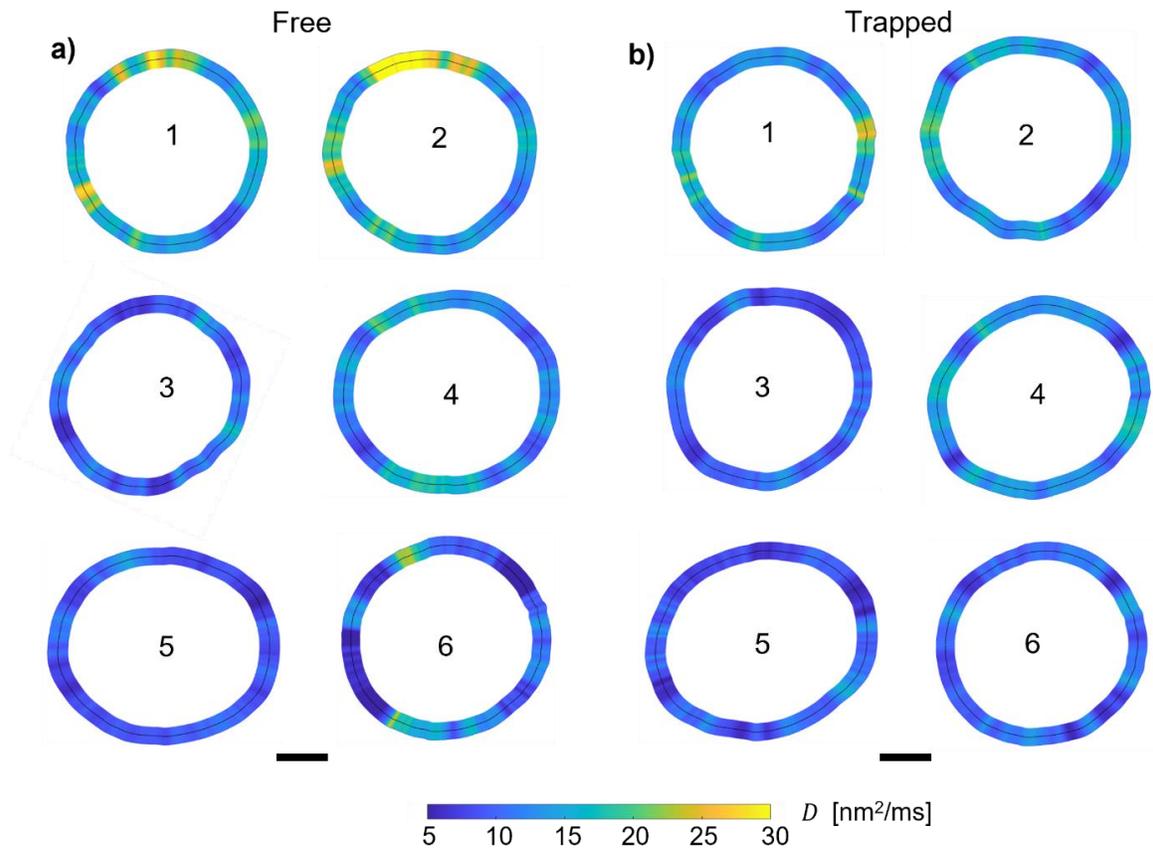

**Fig. S5.** Effective diffusivity maps of single RBCs. a) RBCs free to oscillate are labelled from 1 to 6. b) The same RBCs reported in a) are analysed when subject to a trapping potential distributed along the membrane circle with a power of 10 mW on the sample plane. The circular maps represent the diffusivity of the cell membrane for each angle. Black lines represent the mean position of the membrane. Scale bar is 2 μm.

**Supporting Information References**

1. Helfrich, W. Elastic Properties of Lipid Bilayers: Theory and Possible Experiments. *Zeitschrift fur Naturforsch. - Sect. C J. Biosci.* **28**, 693–703 (1973).
2. Rodríguez-García, R. *et al.* Direct Cytoskeleton Forces Cause Membrane Softening in Red Blood Cells. *Biophys. J.* **108**, 2794–2806 (2015).
3. Betz, T., Lenz, M., Joanny, J. F. & Sykes, C. ATP-dependent mechanics of red blood cells. *Proc. Natl. Acad. Sci. U. S. A.* **106**, 15320–15325 (2009).